# Dual-Sublattice Ferromagnetism Driven by Cooperative Double Exchange and Superexchange at $NdNiO_3/CaMnO_3$ Interfaces

Sharup Sheikh[1], Uditha M. Jayathilake[1], Michael Terilli[2], Mikhail Kareev[2], Jay R. Paudel[3], Arian Arab[1], Christoph Klewe[4], Tien-Lin Lee[5], Jak Chakhalian[2], and Alexander X. Gray[1]

[1]*Department of Physics, Temple University, Philadelphia, Pennsylvania 19122, USA*

[2]*Department of Physics and Astronomy, Rutgers University, Piscataway, New Jersey 08854, USA*

[3]*Chemical Sciences Division, Lawrence Berkeley National Laboratory, Berkeley, California 94720, USA*

[4]*Advanced Light Source, Lawrence Berkeley National Laboratory, Berkeley, California 94720, USA*

[5]*Diamond Light Source Ltd., Didcot, Oxfordshire OX11 0DE, United Kingdom*

* *axgray@temple.edu*

Engineering emergent ferromagnetism at correlated-oxide interfaces offers a powerful route to creating collective states that do not exist in the parent materials. Here, we show that interfacial valence reconstruction in $NdNiO_3/CaMnO_3$ superlattices generates dual-sublattice ferromagnetism involving both Mn and Ni. Depth-resolved standing-wave X-ray photoelectron spectroscopy reveals enhanced $Mn^{3+}$ character on the $CaMnO_3$ side of the interface and enhanced $Ni^{2+}$ character on the $NdNiO_3$ side, establishing the configurations required for $Mn^{4+}$-O-$Mn^{3+}$ double exchange and $Ni^{2+}$-O-$Mn^{4+}$ superexchange, respectively. At low temperature, element-specific XMCD reveals ferromagnetic responses from both sublattices, with the Ni response concentrated predominantly in the $Ni^{2+}$-derived spectral component that SW-XPS independently shows to be enhanced at the interface. Together, these results show that cooperative double exchange within $CaMnO_3$ and superexchange across the interface couple the Mn and Ni sublattices, providing a general strategy for engineering interfacial ferromagnetism in correlated oxides.

Emergent magnetism at correlated-oxide interfaces can arise from electronic configurations that do not exist in either parent material, providing a powerful route for engineering collective states at the atomic scale [1–4]. At these interfaces, charge redistribution, orbital reconstruction, epitaxial strain, symmetry breaking, and oxygen nonstoichiometry can reorganize the local electronic structure within only a few unit cells [5–11]. Because magnetic exchange is highly sensitive to transition-metal valence and metal-oxygen hybridization, such interfacial reconstruction can activate exchange pathways that are absent in the bulk constituents [12,13]. Establishing how these reconstructed valence states vary with depth, and how the resulting exchange interactions couple distinct magnetic sublattices, is therefore essential for understanding and controlling emergent magnetism in oxide heterostructures.

$CaMnO_3$-based heterostructures provide a particularly useful platform for addressing this problem. Bulk $CaMnO_3$ is a G-type antiferromagnetic insulator with Mn predominantly in the $Mn^{4+}$ state [14], whereas electron doping or oxygen-vacancy formation can generate mixed $Mn^{3+}/Mn^{4+}$ configurations and activate ferromagnetic $Mn^{4+}$-O-$Mn^{3+}$ double exchange [3,15]. Rare-earth nickelates provide a complementary correlated component. $NdNiO_3$ exhibits strong coupling among electronic, magnetic, and lattice degrees of freedom and undergoes a temperature-driven metal-insulator transition accompanied by antiferromagnetic ordering [16,17]. Juxtaposing these distinct electronic and magnetic ground states creates an interface at which both Mn and Ni valence configurations, and consequently their exchange interactions, can be reconstructed within only a few unit cells.

Related $LaNiO_3/CaMnO_3$ heterostructures have already demonstrated emergent interfacial ferromagnetism associated with valence reconstruction at nickelate/manganite interfaces [3,4,12,18,19]. Reduction of Mn toward $Mn^{3+}$ can generate $Mn^{4+}$-O-$Mn^{3+}$ double exchange within $CaMnO_3$, while interfacial $Ni^{2+}$ can enable $Ni^{2+}$-O-$Mn^{4+}$ superexchange across the nickelate/manganite boundary. $NdNiO_3$, however, places the nickelate layer in a qualitatively different correlated regime from metallic $LaNiO_3$, raising the question of whether interfacial reconstruction can simultaneously activate both exchange pathways and couple the Mn and Ni sublattices. More fundamentally, the depth-dependent Mn and Ni valence profiles across buried $NdNiO_3/CaMnO_3$ interfaces, and their direct relationship to the magnetic response, remain unresolved. Addressing this problem requires a probe capable of distinguishing interfacial from layer-interior electronic states within individual buried layers.

Here, we use standing-wave X-ray photoelectron spectroscopy (SW-XPS) to resolve the electronic reconstruction across the four-unit-cell-thick $NdNiO_3$ and $CaMnO_3$ layers with unit-cell-scale depth sensitivity, complemented by X-ray absorption spectroscopy (XAS) and X-ray magnetic circular dichroism (XMCD). SW-XPS reveals enhanced $Mn^{3+}$ character on the $CaMnO_3$ side of the interface and an enhanced $Ni^{2+}$ population on the $NdNiO_3$ side, establishing the interfacial valence configurations required for $Mn^{4+}$-O-$Mn^{3+}$ double exchange and $Ni^{2+}$-O-$Mn^{4+}$ superexchange, respectively. At low temperature, XMCD demonstrates a ferromagnetic response from both transition-metal sublattices, with the Ni response concentrated predominantly in the $Ni^{2+}$-derived spectral component that SW-XPS independently identifies as interfacially enhanced. Together, these results establish interfacial dual-sublattice ferromagnetism in which $Mn^{4+}$-O-$Mn^{3+}$ double exchange within $CaMnO_3$ and $Ni^{2+}$-O-$Mn^{4+}$ superexchange across the interface act cooperatively. These findings identify depth-dependent interfacial valence reconstruction as a route to activating multiple exchange channels and coupling distinct magnetic sublattices in correlated-oxide heterostructures.

A high-quality $(NdNiO_3/CaMnO_3)\times10$ superlattice was grown epitaxially on single-crystal $LaAlO_3$ (001) by pulsed-laser interval deposition [20], with each constituent layer nominally four pseudocubic unit cells thick. In situ reflection high-energy electron diffraction (RHEED) confirmed layer-by-layer growth, while ex situ X-ray diffraction (XRD) verified the crystalline quality of the superlattice. Bulk-sensitive hard X-ray photoelectron spectroscopy (HAXPES) confirmed the chemical composition. Additional growth and chemical characterization are provided in Figures S1-S2 of the Supporting Information.

Depth-resolved SW-XPS measurements were performed at the I09 beamline of Diamond Light Source [21] at 296 K. Immediately before the SW-XPS measurements, the Nd $M_{4,5}$-edge XAS spectrum of the same superlattice was recorded to determine the excitation energy that maximized the standing-wave response. A photon energy of 976 eV, on the low-energy side of the Nd $M_5$ ($3d_{5/2}$) resonance, was selected to enhance first-order Bragg reflectivity and standing-wave contrast. The grazing-incidence angle, measured relative to the sample surface, was scanned from 8.8° to 16.8° across the first-order Bragg condition. Photoelectrons were detected with a Scienta EW4000 analyzer at an overall energy resolution of ~300 meV.

The SW-XPS geometry is illustrated in Figure 1a. At the first-order Bragg condition, interference between the incident and reflected X-rays generates a standing-wave electric field whose intensity varies periodically with depth [22]. Scanning the grazing-incidence angle across the Bragg condition shifts the standing-wave phase vertically through the superlattice by half of its period. This controlled displacement of the field maximum provides unit-cell-scale depth selectivity, allowing interfacial and layer-interior contributions to be distinguished within the same buried layer [23].

To establish the structural depth profile and standing-wave response, angle-dependent photoelectron yield curves were measured for representative Ca, Mn, O, Nd, Ni, and C core levels and self-consistently modeled using X-ray optical calculations [24] (Figure 1b). The calculations incorporate element-specific photoelectric cross sections and inelastic mean free paths, with optical constants independently derived from XAS measurements of the same superlattice followed by Kramers-Kronig analysis [25]. The calculated curves reproduce the measured amplitudes, phases, and line shapes across all elements. In particular, the Ca and Mn photoelectron yield curves are strongly phase-shifted relative to the Nd and Ni signals, directly reflecting their distinct vertical positions within the alternating $CaMnO_3$ and $NdNiO_3$ layers.

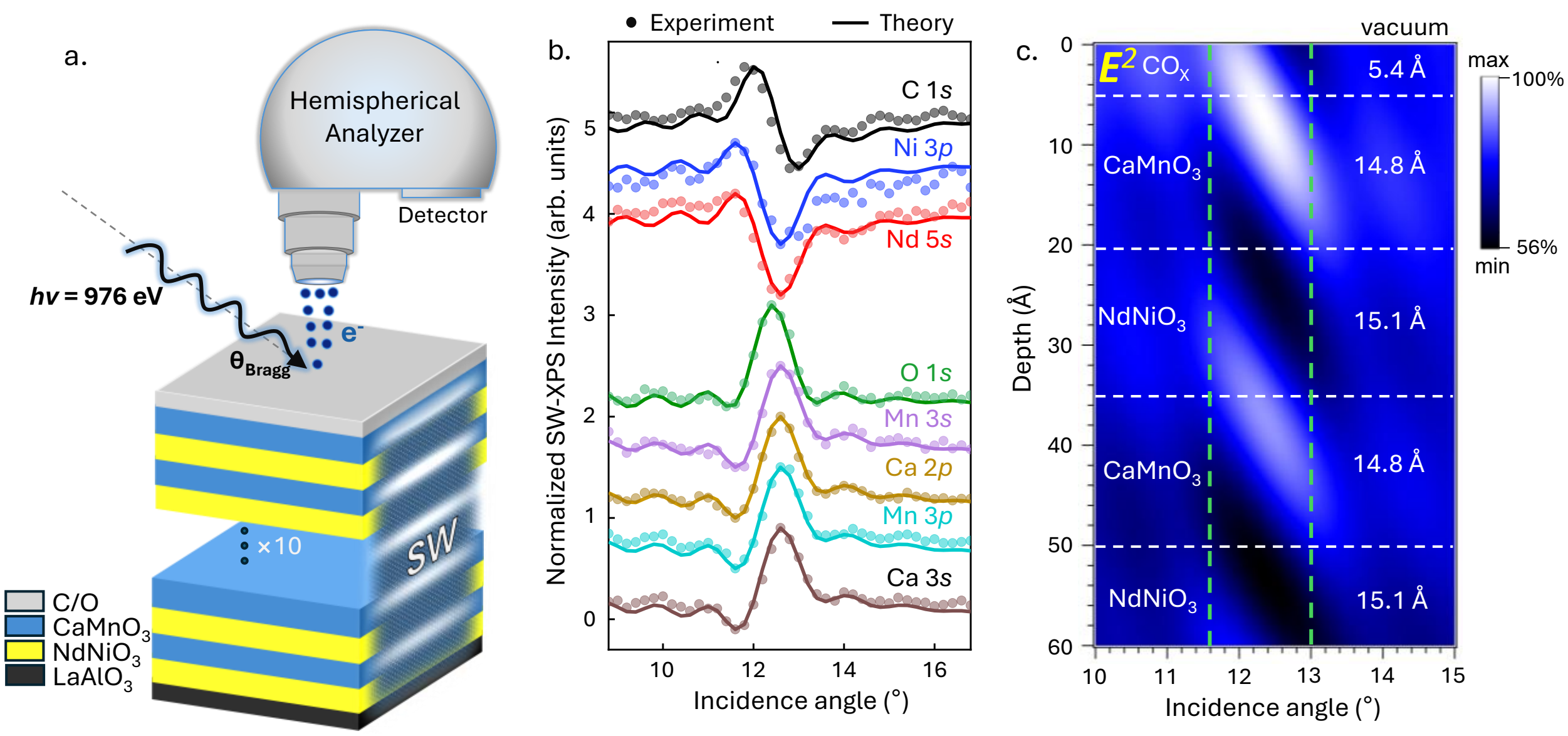

**Figure 1. Soft X-ray standing-wave geometry and depth-selective photoemission response in the $NdNiO_3/CaMnO_3$ superlattice.** (a) Schematic of the SW-XPS geometry for the epitaxial $(NdNiO_3/CaMnO_3)\times10$ superlattice grown on $LaAlO_3(001)$, with each constituent layer nominally four pseudocubic unit cells thick. (b) Experimental core-level photoelectron yield curves (symbols) and corresponding best-fit X-ray optical calculations (solid lines) for representative Ca, Mn, O, Nd, Ni, and C signals. The relative phases and amplitudes of the photoelectron yield curves establish the structural depth profile used in the standing-wave calculations. (c) Calculated standing-wave electric-field intensity, $|E|^2$, at $h\nu = 976$ eV as a function of depth and grazing-incidence angle, measured relative to the sample surface. Scanning the incidence angle across the first-order Bragg condition shifts the standing-wave antinodes through the constituent layers, enabling unit-cell-scale differentiation of interfacial and layer-interior electronic states.

The best-fit model yields $CaMnO_3$ and $NdNiO_3$ layer thicknesses of 14.8 and 15.1 Å, respectively, in close agreement with the nominal four-unit-cell values. A thin surface $CO_x$ layer (~5.4 Å), identified through its C 1*s* signal, was included in the model and provides an independent depth reference. Together, these parameters establish the structural depth scale used to calculate the standing-wave field within the superlattice.

Using this validated structural model, we calculate the standing-wave electric-field distribution shown in Figure 1c. At 11.6°, the standing-wave antinodes are positioned near the $CaMnO_3$ layer boundaries, maximizing sensitivity to interfacial Mn while the layer center lies near a node. Increasing the incidence angle to 13.0° shifts the intensity maximum toward the center of the $CaMnO_3$ layer, enhancing sensitivity to layer-interior Mn while simultaneously increasing sensitivity to interfacial Ni in the adjacent $NdNiO_3$ layer. The standing-wave contrast reaches ~44% between nodal and antinodal positions, allowing the interfacial and layer-interior electronic structures to be distinguished within a single four-unit-cell layer and providing the basis for the element-specific depth-resolved Mn and Ni valence analyses below.

Having established the standing-wave depth selectivity, we next use the Mn 3*s* core level to quantify the Mn valence across the four-unit-cell $CaMnO_3$ layer. Figure 2a shows the calculated standing-wave intensity profiles at 11.6° and 13.0°, which enhance sensitivity to the interfacial and layer-interior regions, respectively. The Mn 3*s* multiplet splitting provides a well-established measure of the average Mn oxidation state because its magnitude decreases systematically with increasing Mn valence [26,27]. Corresponding Mn 3*s* spectra acquired at the two geometries are shown in Figure 2b.

A clear depth dependence of the Mn valence is observed. At the interface-sensitive geometry (11.6°), the Mn 3*s* splitting is 4.93 eV, corresponding to an average formal Mn valence of approximately +3.41 and therefore a substantial $Mn^{3+}$ contribution. At the layer-interior-sensitive geometry (13.0°), the splitting decreases to 4.74 eV, corresponding to a higher average formal Mn valence of approximately +3.64. Thus, Mn is more reduced near the $NdNiO_3$/$CaMnO_3$ interface than toward the interior of the $CaMnO_3$ layer.

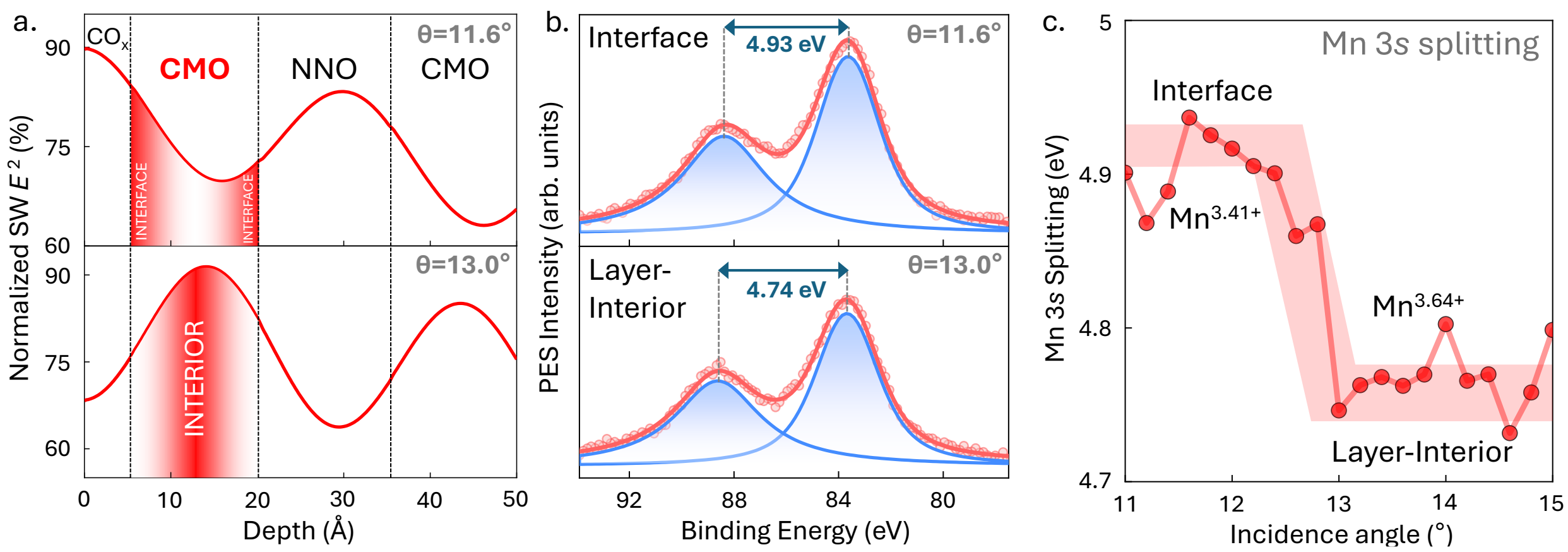

**Figure 2. Depth-resolved Mn valence reconstruction within $CaMnO_3$ at the $NdNiO_3/CaMnO_3$ interface.** (a) Calculated normalized standing-wave electric-field intensity, $|E|^2$, as a function of depth at grazing-incidence angles of 11.6° and 13.0°, which enhance sensitivity to interfacial and layer-interior Mn within the four-unit-cell $CaMnO_3$ layer, respectively. (b) Mn 3*s* core-level spectra acquired at the interface-sensitive (11.6°) and layer-interior-sensitive (13.0°) geometries, together with the corresponding fits used to determine the Mn 3*s* multiplet splitting. The splitting decreases from 4.93 eV at 11.6° to 4.74 eV at 13.0°, corresponding to average formal Mn valences of approximately +3.41 and +3.64, respectively. (c) Mn 3*s* multiplet splitting as a function of grazing-incidence angle across the first-order Bragg condition. The systematic angular dependence reveals enhanced $Mn^{3+}$ character near the interface and a progressively more $Mn^{4+}$-like state toward the $CaMnO_3$ layer interior, corresponding to an interfacial reduction of the average Mn valence by approximately 0.23.

To follow this reconstruction continuously through the layer, the incidence angle was scanned across the Bragg condition in 0.2° steps. The resulting Mn 3*s* splitting (Figure 2c) evolves systematically with the standing-wave position, revealing an enhanced $Mn^{3+}$ population near the interface and a progressively more $Mn^{4+}$-like state toward the layer interior. This corresponds to an interfacial reduction of the average Mn valence by approximately 0.23 relative to the layer interior. Small oscillations at the limits of the angular scan arise from finite-thickness interference (Kiessig fringes) that modulates the standing-wave intensity.

The depth-resolved Mn valence profile therefore provides direct evidence for an interfacial $Mn^{3+}/Mn^{4+}$ mixed-valence state within $CaMnO_3$. This interfacial reconstruction creates the local electronic configuration required for $Mn^{4+}$-O-$Mn^{3+}$ double exchange, identifying the first ferromagnetic exchange pathway underlying the dual-sublattice ferromagnetism discussed below.

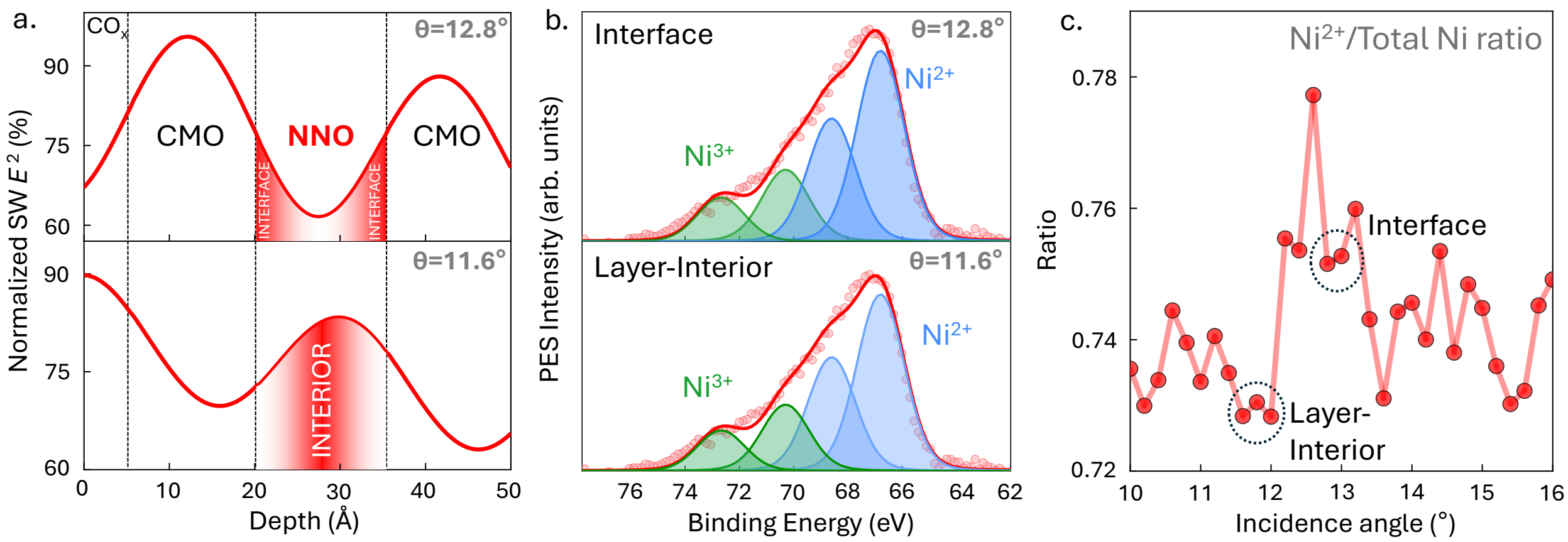

**Figure 3. Depth-resolved Ni valence reconstruction within $NdNiO_3$ at the $NdNiO_3/CaMnO_3$ interface.** (a) Calculated normalized standing-wave electric-field intensity, $|E|^2$, as a function of depth at grazing-incidence angles of 12.8° and 11.6°, which enhance sensitivity to interfacial and layer-interior Ni within the four-unit-cell $NdNiO_3$ layer, respectively. (b) Ni 3*p* core-level spectra acquired at the interface-sensitive (12.8°) and layer-interior-sensitive (11.6°) geometries, together with fits containing $Ni^{2+}$- and $Ni^{3+}$-derived components. (c) $Ni^{2+}$ fraction, defined as the fitted $Ni^{2+}$ component area relative to the total Ni 3*p* spectral area, as a function of grazing-incidence angle. The $Ni^{2+}$ fraction is enhanced under interface-sensitive conditions, reaching ~0.75 compared with ~0.73 in the layer-interior-sensitive geometry, demonstrating an increased $Ni^{2+}$ population at the $NdNiO_3/CaMnO_3$ interface.

Having identified the $Mn^{3+}/Mn^{4+}$ configuration that supports double exchange, we next examine whether $NdNiO_3$ can provide a second exchange pathway by mapping the depth distribution of $Ni^{2+}$, the valence state required for interfacial $Ni^{2+}$-O-$Mn^{4+}$ superexchange [19]. Figure 3a shows the calculated standing-wave intensity profiles at 12.8° and 11.6°, which enhance sensitivity to interfacial and layer-interior Ni, respectively. The corresponding Ni 3*p* spectra (Figure 3b) exhibit mixed $Ni^{2+}/Ni^{3+}$ character and were fitted using $Ni^{2+}$ and $Ni^{3+}$ Voigt components following Shirley background subtraction [4,28,29]. The $Ni^{2+}$ fraction, defined as the fitted $Ni^{2+}$ component area relative to the total Ni 3*p* spectral area (Figure 3c), varies systematically with the standing-wave position and is enhanced under interface-sensitive conditions, reaching ~0.75 compared with ~0.73 in the layer-interior-sensitive geometry. This depth dependence demonstrates an enhanced $Ni^{2+}$ population at the $NdNiO_3/CaMnO_3$ interface, establishing the interfacial electronic configuration required for $Ni^{2+}$-O-$Mn^{4+}$ superexchange. Together with the $Mn^{3+}/Mn^{4+}$ reconstruction established above, these results identify two distinct interfacial ferromagnetic exchange pathways whose consequences are examined below. The interfacial reduction of both transition-metal species may arise from a combination of charge redistribution, oxygen-vacancy-induced electron doping, polar charge compensation, and local bonding reconstruction [3,4,12,18,30–32].

We next use XAS and XMCD to determine whether these two interfacial exchange pathways produce a ferromagnetic response on both the Mn and Ni sublattices. Measurements were performed at beamline 4.0.2 of the Advanced Light Source, Lawrence Berkeley National Laboratory [33]. Spectra were collected at the Mn and Ni $L_{2,3}$ edges in total electron yield (TEY) mode with an energy resolution of ~100 meV at both 295 and 20 K. XMCD spectra were acquired at both temperatures under an in-plane magnetic field of 3.5 T, enabling direct comparison of the room-temperature and low-temperature ferromagnetic responses.

Figures 4a and 4b show the Mn and Ni $L_{2,3}$-edge XAS and XMCD spectra measured at 295 and 20 K. At the Mn $L_3$ edge, features near 640 and 643 eV are associated with $Mn^{3+}$ and $Mn^{4+}$, respectively [32], while the Ni $L_3$ spectrum contains distinct $Ni^{2+}$- and $Ni^{3+}$-derived features near 853 and 854.5 eV [34,35]. These XAS line shapes exhibit no appreciable temperature-dependent change, showing that the overall mixed $Mn^{3+}/Mn^{4+}$ and $Ni^{2+}/Ni^{3+}$ character persists into the low-temperature regime. At 20 K, clear XMCD signals are observed at both the Mn and Ni edges, demonstrating a ferromagnetic response from both transition-metal sublattices. With each XAS spectrum normalized to unity, the XMCD signals reach approximately -0.04 at the Mn $L_3$ edge and

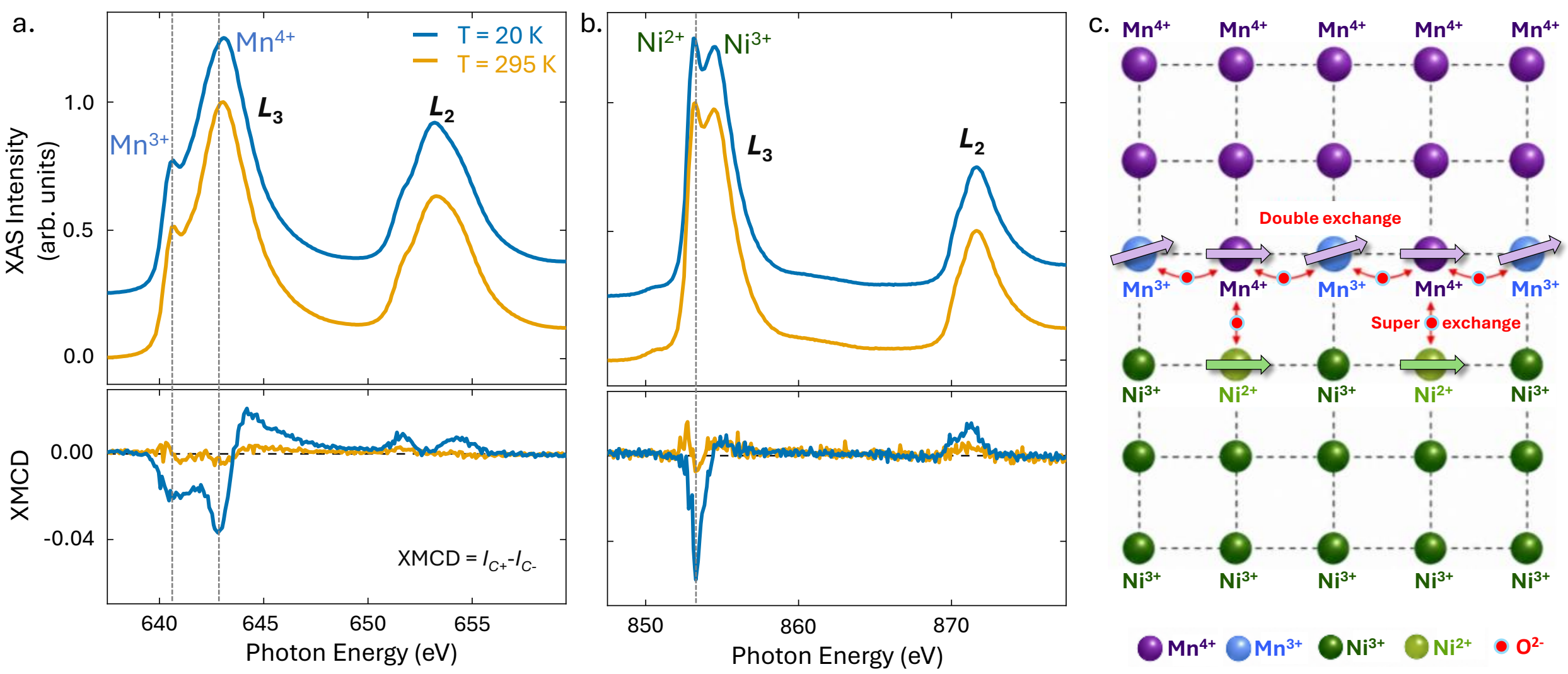

**Figure 4. Dual-sublattice ferromagnetism and cooperative interfacial exchange in the $NdNiO_3/CaMnO_3$ superlattice.** (a) Mn $L_{2,3}$-edge XAS and XMCD spectra measured at 20 and 295 K. The XAS line shape shows no appreciable temperature-dependent change, indicating that the overall mixed $Mn^{3+}/Mn^{4+}$ character persists across this temperature range. At 20 K, the XMCD spectrum exhibits a clear ferromagnetic response at the Mn $L_3$ edge. With the XAS maximum normalized to unity, the XMCD reaches approximately -0.04, corresponding to a dichroic response of ~4%, whereas the 295 K XMCD signal is nearly negligible. (b) Ni $L_{2,3}$-edge XAS and XMCD spectra measured under the same conditions. The XAS line shape likewise shows no significant temperature-dependent change, consistent with persistent mixed $Ni^{2+}/Ni^{3+}$ character. At 20 K, the Ni XMCD exhibits a pronounced ferromagnetic response, reaching approximately -0.05 (~5% of the XAS maximum) and concentrated predominantly in the $Ni^{2+}$-derived spectral component, while the 295 K XMCD signal is nearly absent. (c) Schematic illustration of the interfacial exchange mechanisms inferred from the combined SW-XPS and XMCD results. Interfacial $Mn^{3+}/Mn^{4+}$ mixed valence within $CaMnO_3$ activates $Mn^{4+}$-O-$Mn^{3+}$ double exchange, while enhanced interfacial $Ni^{2+}$ in $NdNiO_3$ enables $Ni^{2+}$-O-$Mn^{4+}$ superexchange across the $NdNiO_3/CaMnO_3$ boundary. Together, these cooperative exchange pathways couple the Mn and Ni sublattices and give rise to interfacial dual-sublattice ferromagnetism.

-0.05 at the Ni $L_3$ edge, corresponding to dichroic responses of approximately 4% and 5% of the respective XAS maxima. In contrast, the XMCD response is negligible at 295 K, showing that the dual-sublattice ferromagnetic response emerges only at low temperature. Importantly, the Ni XMCD response is concentrated predominantly in the $Ni^{2+}$-derived spectral component. Because SW-XPS independently shows that $Ni^{2+}$ is enhanced at the buried interface, the valence-selective XMCD response provides strong evidence that the Ni ferromagnetism is itself interfacial.

Taken together, the depth-resolved SW-XPS and valence-selective XMCD measurements establish a coherent microscopic picture of interfacial dual-sublattice ferromagnetism (Figure 4c). Within $CaMnO_3$, SW-XPS directly resolves an enhanced interfacial $Mn^{3+}$ population, creating the $Mn^{3+}/Mn^{4+}$ mixed-valence configuration required for $Mn^{4+}$-O-$Mn^{3+}$ double exchange [3,36,37]. On the $NdNiO_3$ side, SW-XPS reveals an enhanced interfacial $Ni^{2+}$ population, while the Ni XMCD signal is concentrated predominantly in the $Ni^{2+}$ spectral component, strongly linking the ferromagnetic Ni response to the reconstructed interface. The resulting $Ni^{2+}$-O-$Mn^{4+}$ superexchange [19,38–40] is intrinsically interfacial because it couples Ni and Mn across the $NdNiO_3/CaMnO_3$ boundary. Together, these cooperative interfacial double-exchange and superexchange pathways couple the Mn and Ni sublattices and give rise to the emergent dual-sublattice ferromagnetism.

In summary, depth-resolved SW-XPS and valence-selective XMCD establish interfacial dual-sublattice ferromagnetism in $NdNiO_3/CaMnO_3$ superlattices. Interfacial valence reconstruction simultaneously generates the $Mn^{3+}/Mn^{4+}$ and $Ni^{2+}/Ni^{3+}$ configurations that activate $Mn^{4+}$-O-$Mn^{3+}$ double exchange within $CaMnO_3$ and $Ni^{2+}$-O-$Mn^{4+}$ superexchange across the interface, while the $Ni^{2+}$-selective XMCD response directly links the Ni ferromagnetism to the reconstructed interface. The cooperation of these distinct exchange channels demonstrates how nanoscale control of valence reconstruction can couple different magnetic sublattices and generate ferromagnetism from antiferromagnetic parent materials. More broadly, these results establish interfacial valence reconstruction as a powerful design principle for creating emergent ferromagnetic states in correlated-oxide heterostructures.

## ACKNOWLEDGMENTS

S.S., U.M.J., and A.X.G. acknowledge support from the U.S. Air Force Office of Scientific Research (AFOSR) under Award No. FA9550-23-1-0476. J.C., M.K., and M.T. acknowledge the support by the U.S. Department of Energy, Office of Science, Office of Basic Energy Sciences, under Award No. DE-SC0022160. A.X.G. also gratefully acknowledges support from the Alexander von Humboldt Foundation. The authors also acknowledge the use of an XRD facility supported by the Laboratory for Research on the Structure of Matter and the NSF through the University of Pennsylvania Materials Research Science and Engineering Center (MRSEC) DMR-2309043.

## *Supporting Information for*

# Dual-Sublattice Ferromagnetism Driven by Cooperative Double Exchange and Superexchange at $NdNiO_3/CaMnO_3$ Interfaces

Sharup Sheikh[1], Uditha M. Jayathilake[1], Michael Terilli[2], Mikhail Kareev[2], Jay R. Paudel[3], Arian Arab[1], Christoph Klewe[4], Tien-Lin Lee[5], Jak Chakhalian[2], and Alexander X. Gray[1]

[1]*Department of Physics, Temple University, Philadelphia, Pennsylvania 19122, USA*

[2]*Department of Physics and Astronomy, Rutgers University, Piscataway, New Jersey 08854, USA*

[3]*Chemical Sciences Division, Lawrence Berkeley National Laboratory, Berkeley, California 94720, USA*

[4]*Advanced Light Source, Lawrence Berkeley National Laboratory, Berkeley, California 94720, USA*

[5]*Diamond Light Source Ltd., Didcot, Oxfordshire OX11 0DE, United Kingdom*

* *axgray@temple.edu*

**Figure S1.** ***In situ*** **RHEED characterization of the $NdNiO_3/CaMnO_3$ superlattice**

The surface crystalline quality of the [4 u.c. $NdNiO_3$/4 u.c. $CaMnO_3$]×10 superlattice grown on $LaAlO_3$ (001) was characterized *in situ* by reflection high-energy electron diffraction (RHEED). Figure S1 shows the RHEED pattern acquired at room temperature immediately after growth. The $NdNiO_3$ and $CaMnO_3$ layers were deposited using 74 and 90 laser pulses per unit cell, respectively, yielding a nominal total superlattice thickness of 30.4 nm. The sharp, symmetric vertical streaks surrounding the intense specular reflection are characteristic of a smooth, crystalline surface and indicate that two-dimensional growth was maintained throughout deposition of the superlattice.

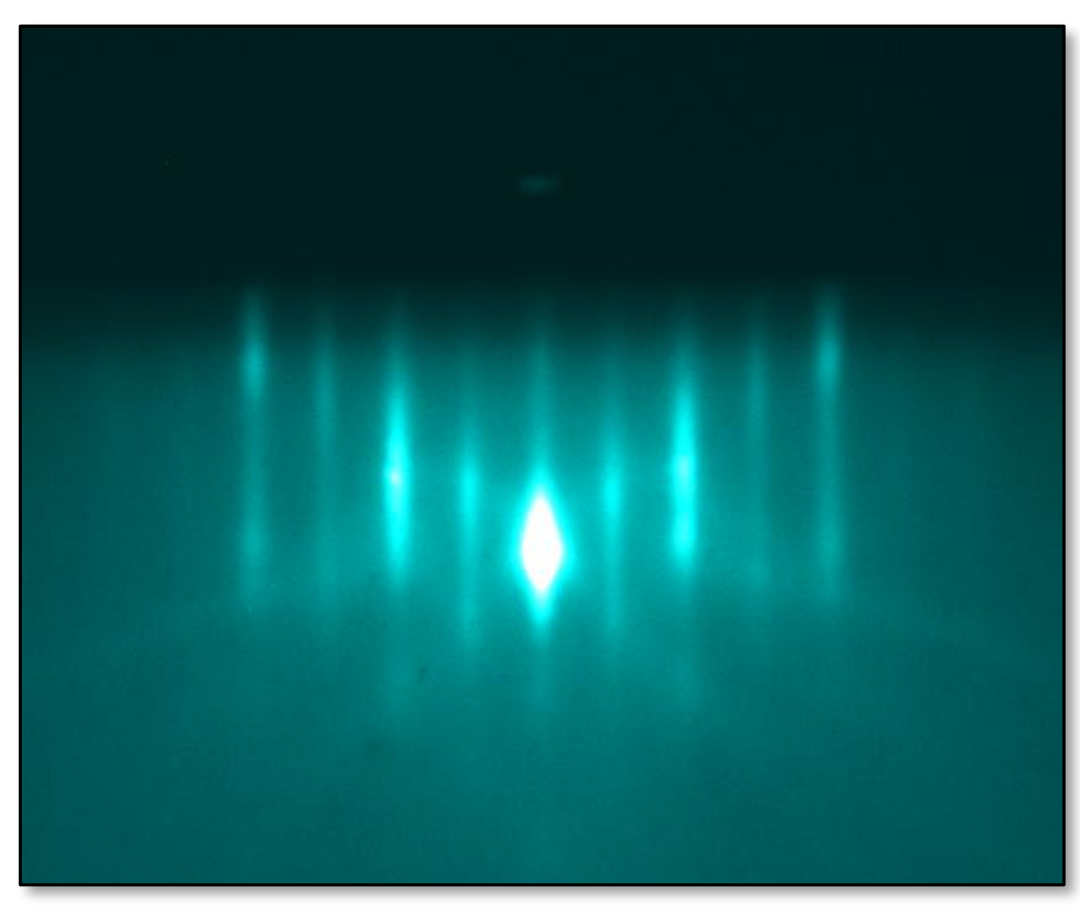

**Figure S1.** Room-temperature RHEED pattern of the [4 u.c. $NdNiO_3$/4 u.c. $CaMnO_3$]×10 superlattice acquired immediately after growth, showing sharp streaks characteristic of a smooth crystalline surface and two-dimensional growth.

**Figure S2. Bulk-sensitive chemical characterization via HAXPES**

To verify the chemical composition of the superlattice, bulk-sensitive hard X-ray photoelectron spectroscopy (HAXPES) was performed at beamline I09 of Diamond Light Source using a photon energy of $hv$ = 6.45 keV and a Scienta Omicron EW4000 hemispherical analyzer. Figure S2 shows the wide-energy-range survey spectrum acquired at room temperature (T = 293 K). Characteristic core-level photoemission features from all expected constituent elements, Nd, Ni, Ca, Mn, and O, are clearly observed, confirming the chemical composition of the superlattice.

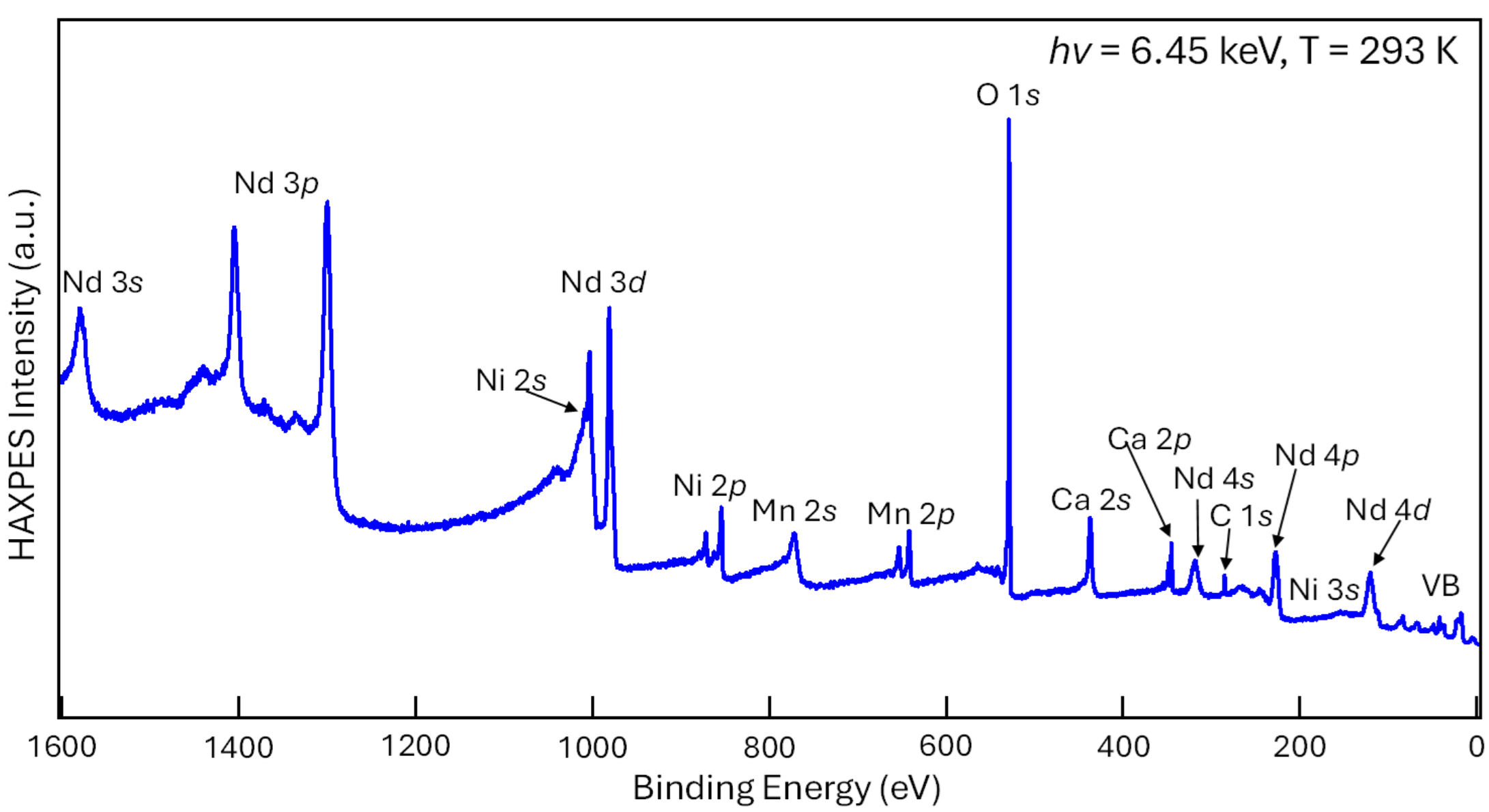

**Figure S2**. HAXPES survey spectrum of the $NdNiO_3/CaMnO_3$ superlattice measured at $hv$ = 6.45 keV, showing the characteristic core-level features of Nd, Ni, Ca, Mn, and O.